\documentclass[prb]{revtex4}

\usepackage{graphicx}
\usepackage{dcolumn}
\usepackage{bm}

\begin{document}

\preprint{APS/123-QED}

\title{Detailed design of a resonantly-enhanced axion-photon regeneration experiment}

\author{Guido Mueller}
\email{mueller@phys.ufl.edu}
\author{Pierre Sikivie}
\author{D.B. Tanner}
\affiliation{%
Department of Physics, University of Florida,\\
Gainesville, FL 32611, USA}

\author{Karl van Bibber}
 \altaffiliation[Also at ]{Lawrence Livermore National Laboratory, Livermore, CA 94550, USA}
\affiliation{Naval Postgraduate School,\\
Monterey, CA 93943, USA}

\date{\today}

\begin{abstract}
A resonantly-enhanced photon-regeneration experiment
to search for the axion or axion-like particles is described. This experiment is a \textit{shining light through walls} study, where photons travelling through a strong magnetic field are (in part) converted to axions; the axions can pass through an opaque wall and convert (in part) back to photons in a second region of strong magnetic field.
The photon regeneration is enhanced by employing matched
Fabry-Perot optical cavities, with one cavity within the axion generation
magnet and the second within the photon regeneration magnet. Compared to simple single-pass photon regeneration, this technique would result in a gain of 
$({\cal F}/\pi)^{2}$, where ${\cal F}$ is the finesse
of each cavity. This gain could feasibly be as high as $10^{10}$, corresponding
to an improvement in the sensitivity to the axion-photon coupling,
$g_{a\gamma\gamma}$, of order $({\cal F}/\pi)^{1/2}\sim 300$.
This improvement would enable, for the first time, a purely laboratory
experiment to probe axion-photon couplings at a level competitive
with, or superior to, limits from stellar evolution or solar axion
searches. This report gives a detailed discussion of the scheme for
actively controlling the two Fabry-Perot cavities and the laser frequencies,
and describes the heterodyne signal detection system, with limits
ultimately imposed by shot noise. 
\end{abstract}

\pacs{PACS numbers: 12.38.-t, 12.38.Qk, 14.80.Mz, 29.90.+r, 95.35.+d}
\maketitle

\section{\label{sec:Intro}Introduction}

The axion remains the most attractive solution to the strong-CP problem
and is also one of two leading dark-matter candidates \cite{Brad03}.
Recently, it has been realized that axions represent a fundamental
underlying feature of string theories; there could be several or even
a great number of axions or axion-like particles within any particular
string theory \cite{Svrc06}.

Present constraints restrict the axion to have very small mass with $mc^{2}$ in the range
between a $\mu$eV and a meV, making the axion extraordinarily weakly coupled
to matter and radiation. Nevertheless, there is considerable ongoing
experimental effort to search for axions. There are at least three
distinct branches to the search. One may detect the axions which constitute
the dark matter, observe axions emitted from the sun's burning core,
or produce and observe axions employing purely laboratory methods.
These last experiments do not depend on cosmological or astrophysical
sources. All of the current efforts rely on the axion's coupling to
two photons; indeed, all are based on the Primakoff effect by which
one of the photons is virtual, whereby an axion can convert into a
single real photon of the same energy (or {\textit{vice versa}})
in a classical electromagnetic field, generally that of a large high-field
superconducting magnet \cite{Siki83}. The coherent mixing of axions
and photons over large spatial regions can compensate for the axion's
extremely weak coupling to a degree sufficient to encourage the experimental
campaigns, although it must be recognized that at the present time,
there is no credible strategy to completely cover the remaining open
parameter space for the axion mass, $m_{a}$, or its coupling to two
photons, $g_{a\gamma\gamma}$.

The simplest and most unambiguous purely laboratory experiment to
look for axions (or light scalars or pseudoscalars more generally)
is photon regeneration (\textit{shining light through the wall}) \cite{vanB87}.
A laser beam traverses a magnetic field, and the field stimulates
a small fraction of photons to convert to axions of the same energy.
A material barrier easily blocks the primary laser beam; in contrast,
the axion component of the beam travels through the wall unimpeded
and enters a second identical arrangement of magnets. There the axions are converted with the same probability back to photons. Because the photon-regeneration
rate goes as $g_{a\gamma\gamma}^{4}$, the sensitivity of the experiment
is poor in its basic form, improved only by increasing the magnetic field strength or
the length of the interaction regions. As initially suggested by Hoogeveen and Ziegenhagen \cite{Hoogeveen91} and briefly presented in a previous letter \cite{Siki07},
very large gains may be realized in both the photon-regeneration rate,
and the resulting limits on $g_{a\gamma\gamma}$ by introducing matched
optical resonators in both the axion production and the photon regeneration magnetic
field regions. In this longer report, we provide a detailed discussion
of an experimental realization, particularly the scheme for locking
two matched high-finesse optical resonators, the signal detection method,
and the ultimate noise limits. Although challenging, the feasibility
of the experiment relies on well-established technologies developed
for example for laser interferometer gravitational-wave detectors \cite{LIGO-NIM}.

Section II will summarize the relevant theory and phenomenology of
the axion. Section III will discuss photon regeneration, and resonantly-enhanced
photon regeneration, presenting experimental results for the former.
Section IV will describe the design and operation of the laser and
optical cavity system, based on a 6+6 string of Fermilab dipole magnets. Section V will present projected results for this specific design.

\section{\label{sec:Axion}The axion in particle physics, astrophysics and cosmology}

The QCD Lagrangian contains a term which violates both $P$ and $CP$,
the so-called \textit{theta term}, 
\begin{equation}
{\cal L}_{\theta}=\frac{\theta g_{s}^{2}}{32\pi^{2}}G_{\mu\nu}^{a}\tilde{G}^{a\mu\nu},\label{Ltheta}
\end{equation}
 where $G_{\mu\nu}^{a}$ is the gluonic field strength, $\tilde{G}^{a\mu\nu}$
is the dual of $G_{\mu\nu}^{a}$, $g_{s}$ is the QCD gauge coupling,
and $\theta$ is a phase angle unprescribed by the Standard Model.
The fact that the strong interactions conserve $P$ and $CP$ to a
very high degree, as evidenced by the upper bound on the neutron electric
dipole moment, implies that $\bar{\theta}\equiv\theta-$ arg (det
$m_{q}$) $<10^{-10}$. There is no reason within the Standard Model why $\bar{\theta}$ should be small. The absence of P and CP violations in the strong interaction constitutes therefore a puzzle, usually referred to as the strong CP problem. 
Peccei and Quinn \cite{PQ} in 1977 discovered an elegant solution
to this problem in which $\theta$ becomes a dynamical
variable of the theory and is driven to its $CP$-conserving value
by the non-perturbative effects which make QCD physics depend upon
$\bar{\theta}$. The axion is the light pseudo-scalar particle which
necessarily results from this theory \cite{WandW}. It has a mass of
\begin{equation}
m_{a}=\frac{f_{\pi}m_{\pi}}{f_{a}}\cdot\frac{\sqrt{m_{u}m_{d}}}{m_{u}+m_{d}}\approx0.6{\rm ~eV}\left(\frac{10^{7}{\rm ~GeV}}{f_{a}}\right)\label{2.1}
\end{equation}
 and a coupling to two photons given by 
\begin{equation}
{\cal L}_{a\gamma\gamma}=g_{\gamma}\cdot\frac{\alpha}{4\pi}\cdot\frac{a}{f_{a}}\cdot F_{\mu\nu}\tilde{F}^{\mu\nu}\ .\label{2.2}
\end{equation}
 Here $f_{a}$ is some large energy scale where $PQ$ symmetry is
spontaneously broken, and $g_{\gamma}$ is a model-dependent parameter
of the theory. In all grand-unified models, for example the Dine-Fischler-Srednicki-Zhitnitskii
model (DFSZ) \cite{DFSZ}, $g_{\gamma}\approx0.36$. In contrast, in
the Kim-Shifman-Vainstein-Zhakarov (KSVZ) model \cite{KSVZ}, $g_{\gamma}\approx-0.97$.
The axion-photon coupling is defined as $g_{a\gamma\gamma}\equiv \alpha |g_{\gamma}|/\pi f_{a}$;
note that thus $g_{a\gamma\gamma}\propto m_{a}$ and limited explorations
of axion models find that for a given mass, $g_{a\gamma\gamma}$ only
varies over about an order of magnitude.

A sufficiently light axion would also be an excellent cold dark matter
(CDM) candidate. The axion mass is bounded from below by a cosmological
constraint, namely the requirement that the present axion cosmological
energy density not exceed the critical density of the universe. From
the vacuum-realignment axion production mechanism the present-day
density is \cite{Vac.} 
\begin{eqnarray}
\rho_{a}^{vac}(t_{0})\approx\rho_{crit}(t_{0})\left(\frac{0.6\cdot10^{-5}{\rm ~eV}}{m_{a}}\right)^{7/6} \nonumber \\
\left(\frac{200{\rm ~MeV}}{\Lambda_{QCD}}\right)^{3/4}
\left(\frac{75{\rm ~km/s}\cdot{\rm Mpc}}{H_{0}}\right)^{2} \label{2.3}
\end{eqnarray}
where $\rho_{crit}\equiv3H_{0}^{2}/8\pi G$ is the present critical
energy density, $H_{0}$ is the present value of the Hubble constant
and $\Lambda_{QCD}$ is the QCD scale factor. Eq.~(\ref{2.3}) implies
the bound $m_{a}>0.6\cdot10^{-5}$~eV.

It should be noted that Eq.~(\ref{2.3}) only provides a rough estimate
of the axion cosmological energy density. Uncertainties result from
many sources; in particular, whether PQ symmetry-breaking occurred
before or after inflation, the nature of the QCD phase transition,
and the possibility of entropy production by late-decaying heavy particles.
Assuming standard concordance cosmology, our estimate of the most likely value
of the axion mass for which $\Omega_{a}=0.22$ is $1.5\cdot10^{-5}$~eV,
if inflation happens before the PQ phase transition. Nevertheless,
given all uncertainties, the lower bound on the axion mass is nominally
taken to be $m_{a}>10^{-6}$~eV.

The mass of the axion is bounded from above by laboratory experiments
and constraints from stellar evolution, the most severe being SN1987a \cite{SN1987a},
which limits the axion mass to $m_{a} < 1.6 \cdot 10^{-2}$~eV. The bound from
SN1987a results from the duration of the neutrino pulse observed by
the IMB and Kamioka detectors (about 20 total, distributed over 10
seconds) and from models of Type-II supernovae. Allowing axions to
be produced from neutron-neutron bremsstrahlung in the core collapse,
i.e. $NN\rightarrow NNa$ would have provided a free-streaming energy
loss channel for axion masses between roughly 0.001 and 1 eV, which
would have foreshortened the neutrino pulse unacceptably.

\begin{figure}
\includegraphics[width=\columnwidth]{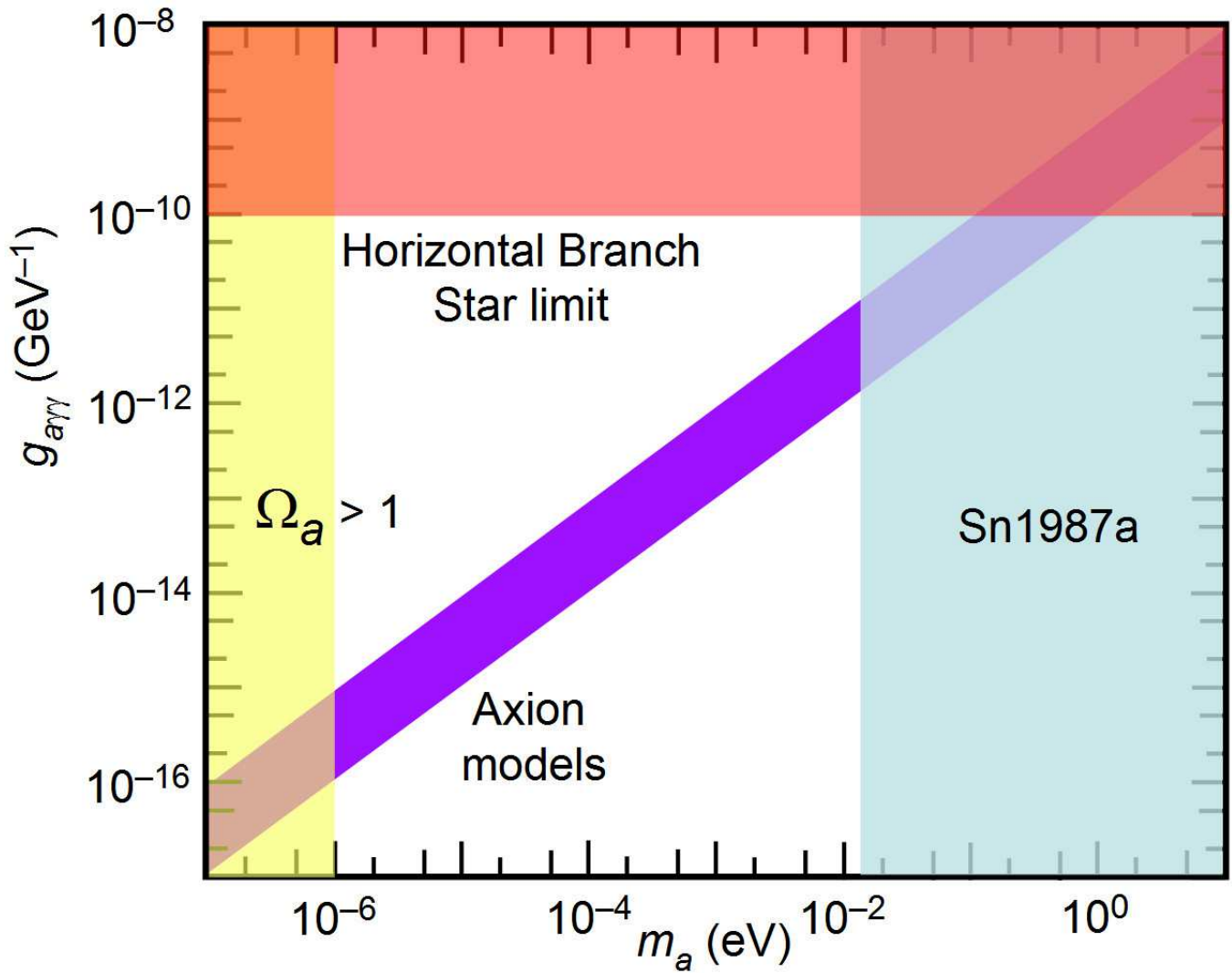}
\caption{\label{fig:axioncon3} Constraints on the axion mass $m_{a}$, and its
coupling to two photons $g_{a\gamma\gamma}$ from cosmology and astrophysics,
along with approximate representation of the band of theoretical models.}
\end{figure}

The axion-photon coupling constant is directly constrained by horizontal
branch (HB) stars to be $g_{a\gamma\gamma}<10^{-10}$~GeV$^{-1}$
\cite{RaffeltBook}. This bound derives from the effect of axions
produced by the Primakoff process in the nuclear burning core, i.e.
$\gamma+Ze\rightarrow a+Ze$, again representing a free-streaming
energy loss channel which would disrupt the agreement of stellar evolution
models with observational systematics of cohort star populations.

Coupled with other stellar bounds and previous laboratory experiments,
all standard axions of mass $m_{a}>10^{-3}$~eV are excluded. Assuming
axions to be the dark matter of our galactic halo, the ADMX microwave
cavity experiment has excluded KSVZ axions within the narrow range
$1.9<m_{a}<3.3$~$\mu$eV; earlier searches set limits up to 16~$\mu$eV
although for considerably stronger couplings \cite{ARNPS}. The CAST
experiment searching for solar axions set limits on $g_{a\gamma\gamma}$
approximately equal to the HB star limit \cite{cast}. The cosmological
and astrophysical limits on the axion, and an approximate representation
of the theoretical models ($m_{a}$ vs. $g_{a\gamma\gamma}$) are
shown in Fig.~\ref{fig:axioncon3}.

\section{\label{sec:resenh}Resonantly enhanced photon regeneration}

The action density for the dynamics of photons and axions is 
\begin{eqnarray}
{\cal L}&=&\frac{1}{2}(\epsilon E^{2}-B^{2})+\frac{1}{2}(\partial_{t}a)^{2}\nonumber \\
&-&\frac{1}{2}(\vec{\nabla}a)^{2}-\frac{1}{2}m_{a}^{2}a^{2}-g_{a\gamma\gamma} a\vec{E}\cdot\vec{B},\label{actden}
\end{eqnarray}
 where $\vec{E}$, $\vec{B}$ and $a$ are respectively the electric,
magnetic and axion fields. The electromagnetic fields are given in
terms of scalar and vector potentials, $\vec{E}=-\vec{\nabla}\Phi-\partial_{t}\vec{A},~\vec{B}=\vec{\nabla}\times\vec{A}$,
as usual. The dielectric function $\epsilon$ is assumed constant
in both space and time. In the presence of a large static magnetic
field $\vec{B}_{0}(\vec{x})$, the equations of motion are 
\begin{eqnarray}
\epsilon\vec{\nabla}\cdot\vec{E} & = & g_{a\gamma\gamma}\vec{B}_{0}\cdot\vec{\nabla}a\nonumber \\
\vec{\nabla}\times\vec{B}-\epsilon\partial_{t}\vec{E} & = & -g_{a\gamma\gamma}\vec{B}_{0}\partial_{t}a\nonumber \\
\partial_{t}^{2}a-\vec{\nabla}^{2}a+m_{a}^{2}a & = & -g_{a\gamma\gamma}\vec{E}\cdot\vec{B_{0}}.\label{eom}
\end{eqnarray}
 $\vec{B}$ now represents the magnetic field minus $\vec{B}_{0}$,
and terms of order $g_{a\gamma\gamma}aE$ and $g_{a\gamma\gamma}aB$ are neglected. Eqs.~(\ref{eom})
describe the conversion of axions to photons and vice-versa.

Using these equations, it can be shown \cite{Siki83,vanB87,Raff88}
that the photon to axion conversion probability $P$ in a region of
length $L$, permeated by a constant magnetic field $B_{0}$ transverse
to the direction of propagation, is given by ($\hbar=c=1$) 
\begin{equation}
P=\frac{1}{4}{\frac{1}{\beta_{a}\sqrt{\epsilon}}}\,(g_{a\gamma\gamma}B_{0}L)^{2}\left(\frac{2}{qL}\sin\frac{qL}{2}\right)^{2},\label{prob}
\end{equation}
where $\beta_{a}$ is the axion speed and $q=k_{a}-k_{\gamma}$ is
the momentum transfer. In terms of the energy $\omega$, which is
the same for the axion and the photon, $k_{a}=\sqrt{\omega^{2}-m_{a}^{2}}$,
$\beta_{a}={k_{a}}/{\omega}$ and $k_{\gamma}=\sqrt{\epsilon}\omega$.
The axion to photon conversion probability in this same region is
also equal to $P$. Everything else being the same, the conversion
probability is largest when $q\approx0$. For $m_{a}<<\omega$, and propagation in a vacuum ($\epsilon=1$), 
\begin{equation}
q=-\frac{m_{a}^{2}}{2\omega}.\label{q}
\end{equation}

Fig.~2a shows the photon regeneration experiment as usually conceived.
If $E_{0}$ is the amplitude of the laser field propagating to the
right, the amplitude of the axion field traversing the wall is $E_{0}\sqrt{P}$
where $P$ is the conversion probability in the magnet on the LHS
of Fig.~1a. Let $P^{\prime}$ be the conversion probability in the
magnet on the RHS. The field generated on that side is then $E_{S}=E_{0}\sqrt{P'P}$
and the number of regenerated photons is $N_{S}=P^{\prime}PN_{0}$
where $N_{0}$ is the number of photons in the initial laser beam.

Fig.~2b shows the two improvements \cite{Hoogeveen91,Siki07} we propose for the experiment.
The first is to build up the electric
field on the left hand side of the experiment using a Fabry-Perot
cavity, as illustrated. We will call this cavity the axion generation cavity.
Assuming an amplitude transmissivity of $t_{1a}$ of the input mirror,
amplitude reflectivities of the two cavity mirrors $r_{1a}$ and $r_{2a}\approx1$,
and a cavity length such that the multiple reflections between the
mirrors interfere constructively, then the right-propagating field
inside the cavity, $E_{a}$, will be: 
\begin{equation}
E_{a}=\frac{2t_{1a}}{t_{1a}^{2}+V_{a}}E_{in}\label{E_a}
\end{equation}
where $E_{in}$ is the amplitude of the laser field going into the
cavity and $V_{a}$ is the fractional power loss the light encounters
during one round-trip inside the cavity, less the losses due to the
transmissivity of the input mirror. Hence, $V_{a}$ includes power
absorption in both mirrors, scattering due to mirror defects, diffraction
from the finite mirror size, and the (small) leakage through mirror
2. The circulating light in the cavity creates an axion field of amplitude
\begin{equation}
a=\sqrt{P}E_{a}=\sqrt{P}\frac{2t_{1a}}{t_{1a}^{2}+V_{a}}E_{in}\label{axionfield}
\end{equation}
As long as $\omega \gg m_a$, the spatial distribution of the axion field is identical to the spatial distribution of the electric field. Assuming the lasers in Fig.~1a and Fig.~1b have the same power, the flux is increased by the
factor $4t_{a}^{2}/(t_{1a}^{2}+V_{a})^{2}$ compared to the case without generation cavity. These axions propagate
through the {}``wall'' and reconvert into photons in the regeneration
region.

\begin{figure}
\includegraphics[width=\columnwidth]{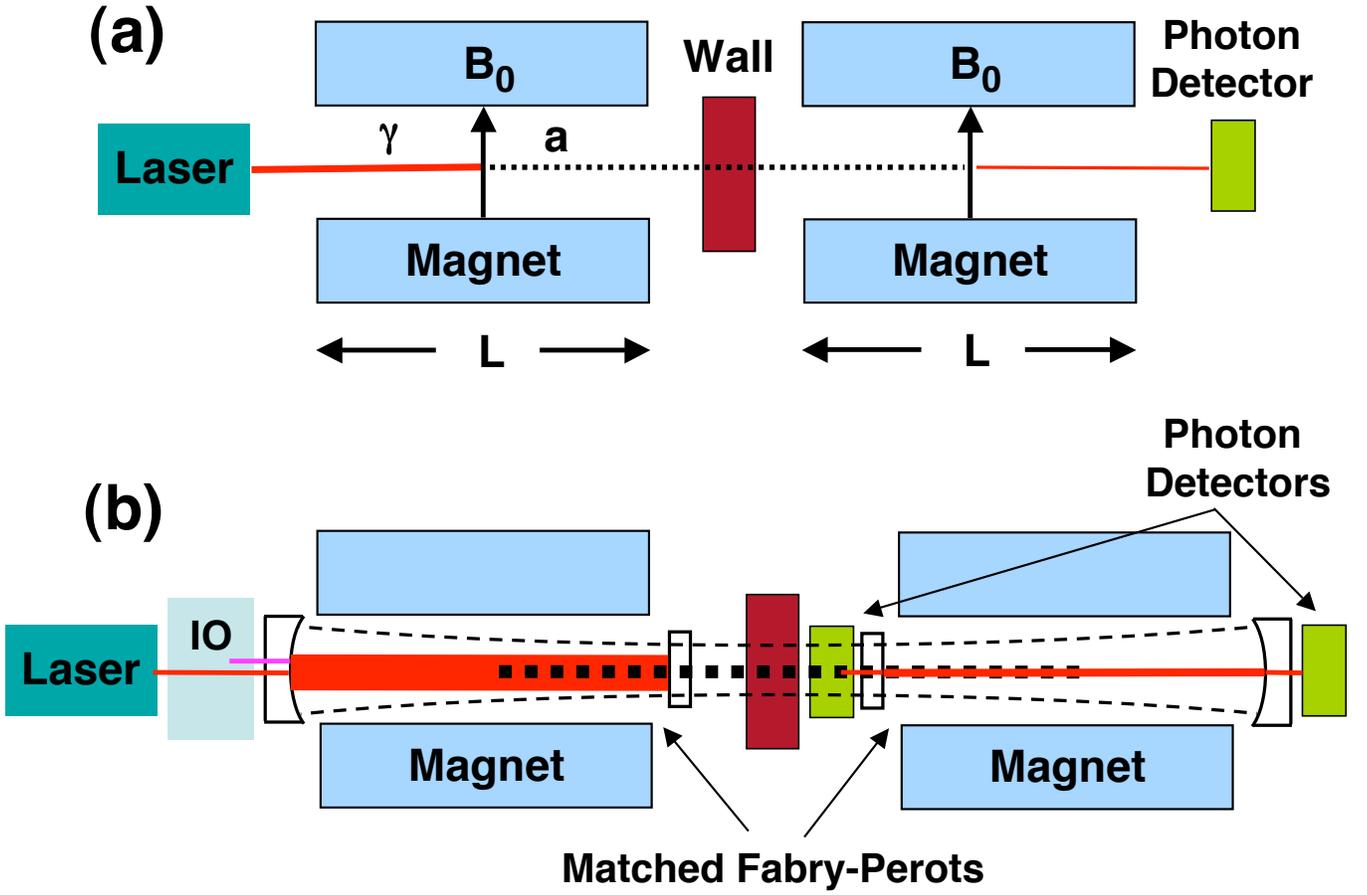} 
\caption{\label{fig:Reprfig1} (a) Simple photon regeneration. (b) Resonant photon regeneration, employing matched Fabry-Perot cavities. The overall envelope schematically
shown by the thin dashed lines indicates the important condition that
the axion wave, and thus the Fabry-Perot mode, in the photon regeneration cavity
must follow that of the hypothetically unimpeded photon wave from
the Fabry-Perot mode in the axion generation magnet. Between the laser and
the cavity is the injection optics (IO) which manages mode matching
of the laser to the cavity, imposes RF sidebands for reflection locking
of the laser to the cavity, and provides isolation for the laser.
The photon detectors are also preceded by matching and beam-steering
optics. Layout is only schematic; as will be seen later, IO will likely be placed between the cavities.}
\end{figure}

As we have said, increasing the axion production rate, and thus the
photon regeneration rate, by building up the optical power in the
first magnet is not a new idea. In fact, the very first photon regeneration
experiment performed and published by the BFRT collaboration \cite{Ruos92,Came93}
utilized an ``optical delay line,'' which caused the laser beam
to traverse the magnet 200 times before exiting. With relatively modest
magnets (4.4 m, 3.7 T each), a limit of $g_{a\gamma\gamma}<6.7\times10^{-7}~{\rm GeV}^{-1}$
was set. Four more photon-regeneration results have recently been reported.
The BMV collaboration utilized a short, pulsed high-field magnet (0.37
m, 12.3 T), and pulsed laser fields ($\omega$ = 1.17 eV, 1.5 kJ/pulse,
4.8 ns, firing every 2 hours), reaching a limit of $g_{a\gamma\gamma}<1.25\times10^{-6}~{\rm GeV}^{-1}$
~ \cite{Robi07}. The GammeV collaboration built upon a single Tevatron
dipole (6m, 5T) with a movable optical barrier in the middle to divide
the magnet into production and regeneration regions, was the most
sensitive, setting an upper bound to the axion-photon coupling of
$g_{a\gamma\gamma}<3.5\times10^{-7}~{\rm GeV}^{-1}$ ~ \cite{Chou08}.
The LIPSS \cite{LIPPS} collaboration used a pulsed free electron laser and two identical magnets (1.77 T, 1 m) and reached a limit similar to the BMV limit. Also the OSQAR experiment \cite{OSQAR} reached a similar upper limit using a 18 W argon laser and one LHC dipole magnet (9.5 T, 14.3 m)separated into two halves by an optical barrier. These limits are valid only for axion masses below about one meV,
with the full sensitivity of the BMV experiment extending
to $m_{a}\approx2$~meV, due to the shorter length of the magnet;
see Figure 3. Other experiments are in various stages of preparation \cite{Ringwald:2006rf}.

%

\begin{figure}
\includegraphics[width=\columnwidth]{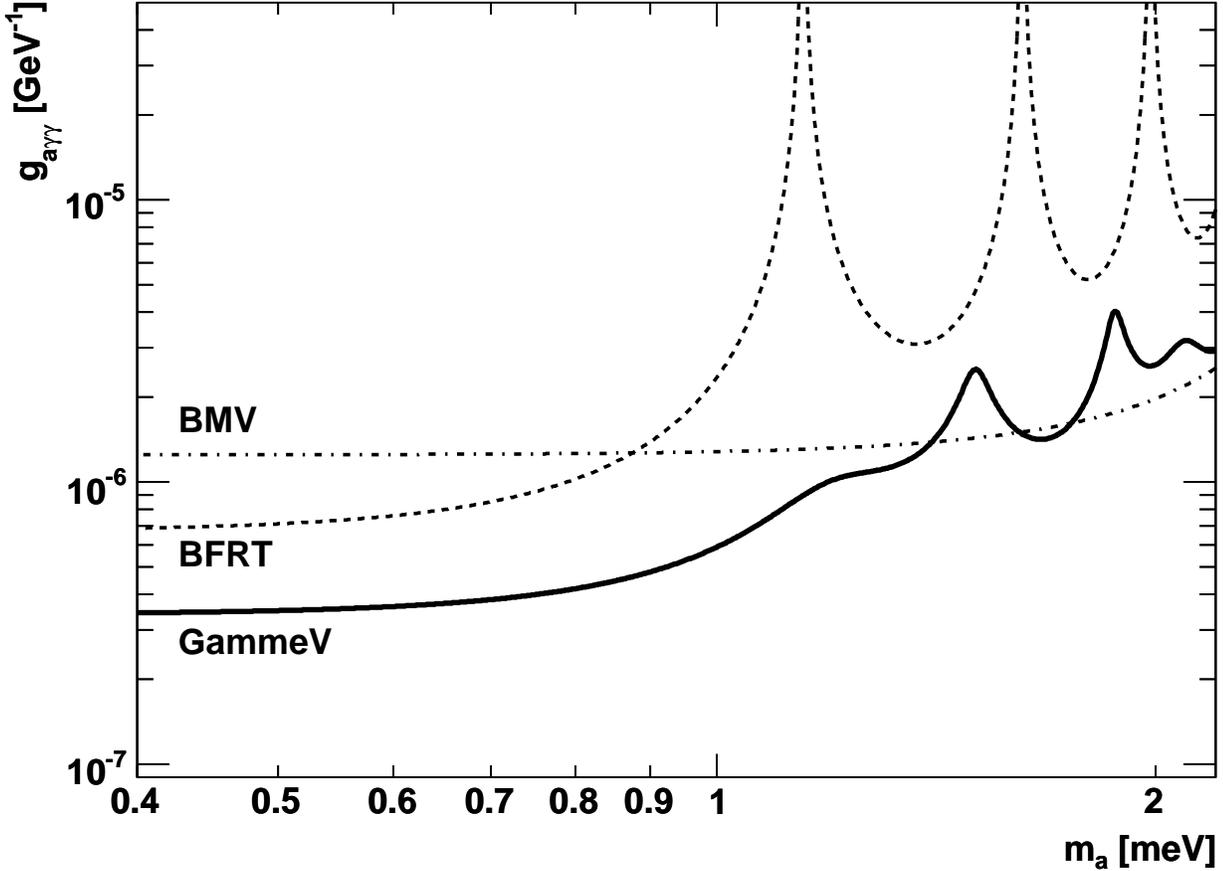} 
\caption{\label{gammev}Exclusion region from three photon-regeneration experiments reported to date.\cite{Ruos92,Came93,Robi07,Chou08} See text for discussion.}
\end{figure}

There is substantial gain from building up the laser power in the
axion production magnet; however, it is immaterial whether one {}``recycles''
the photons incoherently, as in an optical delay line, or coherently,
as in a Fabry-Perot cavity. In contrast, the coherent case alone can
provide a large additional gain in sensitivity for photon regeneration.
Thus, the second improvement \cite{Hoogeveen91,Siki07} is to install also a second Fabry-Perot
cavity, the photon regeneration cavity, on the other side of the experiment,
making a symmetric arrangement, as illustrated on the right-hand side
of Fig.~2b. In this setup, the axion field acts as a source field
similar to a gain medium in a laser resonator. The intra-cavity field
can be calculated using the equilibrium condition: 
\begin{eqnarray}
E_{\gamma}&=&\frac{1}{1-r_{1\gamma}r_{2\gamma}e^{i\phi_{RT}}}\eta\sqrt{P}ae^{ik_{a}d}\label{E_c} 
\end{eqnarray}
 where $\phi_{RT}$ is the round trip phase of the field, $d$ is
the distance between the two cavities and $r_{1\gamma}$ and $r_{2\gamma}$
are the amplitude reflectivities of the two cavity mirrors. $\eta$
is the spatial overlap integral between the axion mode and the electric
field mode. This overlap will be identical to unity (up to corrections of order
$m_{a}/\omega_{0}$) if the spatial eigenmodes of the two cavities
are extensions of each other, e.g. when the Gaussian eigenmode in
one cavity propagated to the other cavity is identical to the Gaussian eigenmode of that cavity.

The field will be resonantly enhanced if $\phi_{RT}=N2\pi$. To detect
the regenerated field, a small part is allowed to transmit through one of the
cavity mirrors, say mirror 1, with an amplitude transmissivity
of $t_{1\gamma}$: 
\begin{equation}
E_{S}=t_{1\gamma}E_{\gamma}\approx\frac{2t_{1\gamma}}{t_{1\gamma}^{2}+V_{\gamma}}\eta\sqrt{P}ae^{ik_{a}d}\label{eq:E_S out}
\end{equation}
Here we replaced the amplitude reflectivities with the
amplitude transmissivities and intensity losses of the mirrors: 
\begin{eqnarray}
r_{1\gamma}=\sqrt{1-t_{1\gamma}^{2}-V_{1\gamma}}\approx1-\frac{t_{1\gamma}^{2}}{2}-\frac{V_{1\gamma}}{2} \nonumber \\ r_{2\gamma}=\sqrt{1-V_{2\gamma}}\approx1-\frac{V_{2\gamma}}{2}\qquad V_{\gamma}\equiv V_{1\gamma}+V_{2\gamma}\label{eq: losses}
\end{eqnarray}
 The final regenerated electric field is: 
\begin{eqnarray}
E_{S}&=&\left(\frac{2t_{1\gamma}}{t_{1\gamma}^{2}+V_{\gamma}}\right)\eta PE_{a}e^{ik_{a}d}\nonumber \\ &=&\left(\frac{2t_{1\gamma}}{t_{1\gamma}^{2}+V_{\gamma}}\right)\left(\frac{2t_{1a}}{t_{1a}^{2}+V_{a}}\right)\eta PE_{in}e^{ik_{a}d}\label{eq:E_s losses}
\end{eqnarray}
 Or in terms of regenerated photons: 
\begin{equation}
N_{S}=\left(\frac{4T_{1\gamma}}{\left(T_{1\gamma}+V_{\gamma}\right)^{2}}\right)\left(\frac{4T_{1a}}{\left(T_{1a}+V_{a}\right)^{2}}\right)\eta^{2}P^{2}N_{in}\label{eq:N_s losses}
\end{equation}
 where we have replaced the amplitude transmissivities with the commonly
used intensity or power transmissivities $T=|t|^{2}$.

The final signal depends on the laser power build-up in the axion
generation cavity and the signal build-up in the photon regeneration cavity.
The build-up in the regeneration cavity is only limited by the losses
$V_{\gamma}$ and the transmissivity $T_{1\gamma}$. As always, the field outside
the cavity is maximum when the cavity is impedance matched ($T=V$).
The losses include coating absorption and scattering from imperfections
in the polished surface, mainly small angle scattering. Current state-of-the-art
mirrors have coating absorption of $<1\,\mbox{ppm}$ and scatter of
$<5\,\mbox{ppm}$ or total losses in the order of $V\approx10\,\mbox{ppm}$
for both mirrors combined. For the impedance-matched case, it is convenient
to express Eq.~(\ref{eq:N_s losses}) in terms of the finesse ${\cal F}_{\gamma,a}$ of the cavities:
\begin{equation}
N_{S}=\eta^{2}\frac{{\cal F}_{\gamma}}{\pi}\frac{{\cal F}_{a}}{\pi}P^{2}N_{in}
\end{equation}
Note that resonant regeneration gives an enhancement factor of $\sim({\cal F}/\pi)^{2}$
over simple photon regeneration. This factor may feasibly be $10^{10}$,
corresponding to an improvement in sensitivity to $g_{a\gamma\gamma}$
of $\approx 300$.

The power build up in the axion generation cavity is further limited by
thermal heating of the mirror surfaces caused by the absorption of
the stored light in the mirror coatings. This heating will change
the radii of curvature of the cavity mirrors, increase the scatter
losses due to non-spherical higher order figure distortions, and
create a thermal lens in the substrate of the input mirror \cite{HelloVinet}.
Current estimates based on Ref. \cite{HelloVinet} suggest that we
can operate the cavity with approximately $100\,\mbox{kW}$ intra-cavity
power without a dedicated thermal correction system. This value could
be increased by at least one order of magnitude using a LIGO-like
thermal correction system, mirror substrates with a higher thermal conductivity then fused silica such as sapphire, and/or cooling the mirrors to lower temperatures
where there is a significant increase in the thermal conductivity
of the mirror substrates \cite{Lawrence}.
The axion generation cavity
can potentially increase the number of generated axions by 5 orders
of magnitude while the photon regeneration cavity potentially increases
the number of regenerated photons by another 5 orders of magnitude.

\section{\label{expimp}Experimental implementation}

The resonantly-enhanced photon regeneration experiment, involving
the design and active locking of high-finesse Fabry-Perot resonators
and the heterodyne detection of weak signals at the shot-noise limit,
is well supported by the laser and optics technology developed for
LIGO \cite{LIGO-NIM}. This section will discuss all aspects of the experimental implementation
while the next section will present estimates of the achievable experimental sensitivity for
a realistic design in terms of power-handling capabilities and mirror
technologies.

\subsection{\label{magnets}Magnets }

The design utilizes a total of 12 Tevatron superconducting dipoles
(each 5T field, 6m length, and 48mm diameter warm bore), 6 for the
axion generation cavity and 6 for the photon regeneration cavity; hereafter referred
to as the {}``TeV 6+6'' configuration. There is a large infrastructure
and experience base for the Tevatron dipoles, and an adequate number
are available.

\subsection{\label{cavitydes}Cavity design}

The layout requires that the optical cavities are mode-matched to
each other. In addition, losses due to aperture effects should be
kept very low. These considerations, together with the dimensions
of the available magnets, drive the design of the two optical cavities.
The waist of the optical eigenmode of both cavities should be half way between the cavities,
at or near the location of the beam block. The magnets plus the necessary
space between each magnet and at the ends of the magnets set the final
length of each cavity. Although all details have not yet been worked
out, we expect that the end mirror of each cavity will be about $46\,\mbox{m}$
away from the waist. The central mirrors can be placed within $2\,\mbox{m}$
of each other or $1\,\mbox{m}$ away from the waist implying a cavity
length of $L_{cav}=45\,\mbox{m}$.

Although Fig. 2 schematically shows the laser light being fed to the axion generation
cavity at the far end (relative to the waist) it is clearly preferable
to inject the light in the space between the mirrors. This approach
minimizes the separation of the injection/locking circuits from the
optics and electronics that lock the photon regeneration cavity. (See discussion
below.) The rest of the paper uses this approach as the baseline.

The free spectral range of each cavity would then be: 
\begin{equation}
FSR=\frac{c}{2\times L_{cav}}\approx3.3\,\mbox{MHz}\label{eq:fsr1}
\end{equation}
The optical cavities have to support mirror-image fundamental spatial
modes and should suppress all higher order spatial modes. Therefore, we choose
$g$-parameters \cite{Siegman} for the optical cavities such that $g_{1}g_{2}=0.6$,
where $g_{i}=1-L_{cav}/R_{i}$, with $R_{i}$ the radii of curvature
of the mirrors. The transversal mode spacing is then: 
\begin{equation}
\Delta\nu_{TEM}=\frac{\mbox{acos}(\sqrt{g_{1}g_{2})}}{\pi}\times FSR\approx726\,\mbox{kHz}\label{eq:tem_LHC}
\end{equation}
Using the dependence of the mirror locations with respect to the
waist of the eigenmode on the $g$-factors and the length of the cavity
\cite{Siegman}, 
\begin{eqnarray}
z_{1}&=&\frac{g_{2}(1-g_{1})}{g_{1}+g_{2}-2g_{1}g_{2}}L_{cav}=-1\,\mbox{m}\nonumber \\ 
z_{2}&=&\frac{g_{1}(1-g_{2})}{g_{1}+g_{2}-2g_{1}g_{2}}L_{cav}=46\,\mbox{m}\label{eq:z_1 z_2}
\end{eqnarray}
we can calculate the individual $g$-factors: 
\begin{equation}
g_{1}=1.015\qquad g_{2}=0.591.\label{eq:g1g2}
\end{equation}
The beam sizes $w_i$ on the mirrors and the waist $w_{0}$
are then: 
\begin{equation}
w_{1}=4.29\,\mbox{mm}\approx w_{0}\qquad w_{2}=5.62\,\mbox{mm}\label{eq:w1w2}
\end{equation}
The beam size at the input mirror is essentially identical to the waist
size. The mirror radii of curvatures are nominally: 
\begin{eqnarray}
ROC_{1}&=&\frac{L_{cav}}{1-g_{1}}=-2946\,\mbox{m}\approx\infty \nonumber \\ 
ROC_{2}&=&\frac{L_{cav}}{1-g_{2}}=110\,\mbox{m}.\label{eq:r1r2}
\end{eqnarray}
If we replace the input mirror by a flat mirror, the loss from mode
mismatch between the two cavities is less than $10^{-4}$ in field
amplitude while the beam sizes would increase by $\sim10\,\mu\mbox{m}$.
For all practical purposes, it is sufficient to assume a flat input
mirror. Note that the intensity on the mirror in the center of the
beam for an intra-cavity power of $1\,\mbox{MW}$ would be: 
\begin{equation}
I=2\frac{P}{\pi w^{2}}\approx3.5\frac{\mbox{MW}}{\mbox{cm}^{2}}\frac{P_{cav}}{[1\,\mbox{MW}]}
\label{eq:intensity}
\end{equation}
well below the damage threshold of modern coatings.

The Tevatron magnets have a $48\,\mbox{mm}$ cold bore and are slightly
curved. They will be attached to each other such that the clear aperture
wiggles around the optical axis but possesses no net bend over all 12 magnets.
We estimated that the clear aperture through the entire 12 magnets
will be $\sim39\,\mbox{mm}$ well beyond the necessary aperture given
the above derived beam sizes. However, clipping at the aperture of
the Tevatron magnets could limit future expansions of the experiment
to about $100\,\mbox{m}$ length for each cavity. 

In addition to mode mismatch because of mismatched beam sizes and
radii of curvatures, the efficiency of photon regeneration will
also depend on alignment mismatches between the cavities. Angular
or lateral misalignment of the optical axes would significantly reduce
the spatial overlap between the two modes. Losses caused by a lateral
shift $\delta x$ scale with the beam size, those caused by an angular shift
scale with the divergence angle, so that the overall efficiency
is: 
\begin{equation}
\eta\approx1-\frac{1}{2}\frac{\delta x^{2}}{w_{1}^{2}}-\frac{1}{2}\frac{\delta\alpha^{2}}{\Theta^{2}}\qquad\mbox{with}:\qquad\Theta=\frac{\lambda}{\pi w_{1}}
\label{eq:misaligment}
\end{equation}
Recall that $\eta$ is the spatial overlap integral between the axion mode and the electric field mode.
Requiring an efficiency $\eta>0.95$, the
requirements on the lateral and angular offsets for the proposed TeV
6+6 configuration are: 
\begin{equation}
\delta x<\sqrt{0.05}w_{1}\approx1\,\mbox{mm};
\end{equation}
\begin{equation}
\delta\alpha<\sqrt{0.05}\Theta\approx17\,\mu\mbox{rad},
\end{equation}
This alignment has to be checked periodically as part of the run
protocol.

\subsection{\label{lenfreq}Length and frequency stabilization system}

\begin{figure}
\includegraphics[width=\columnwidth]{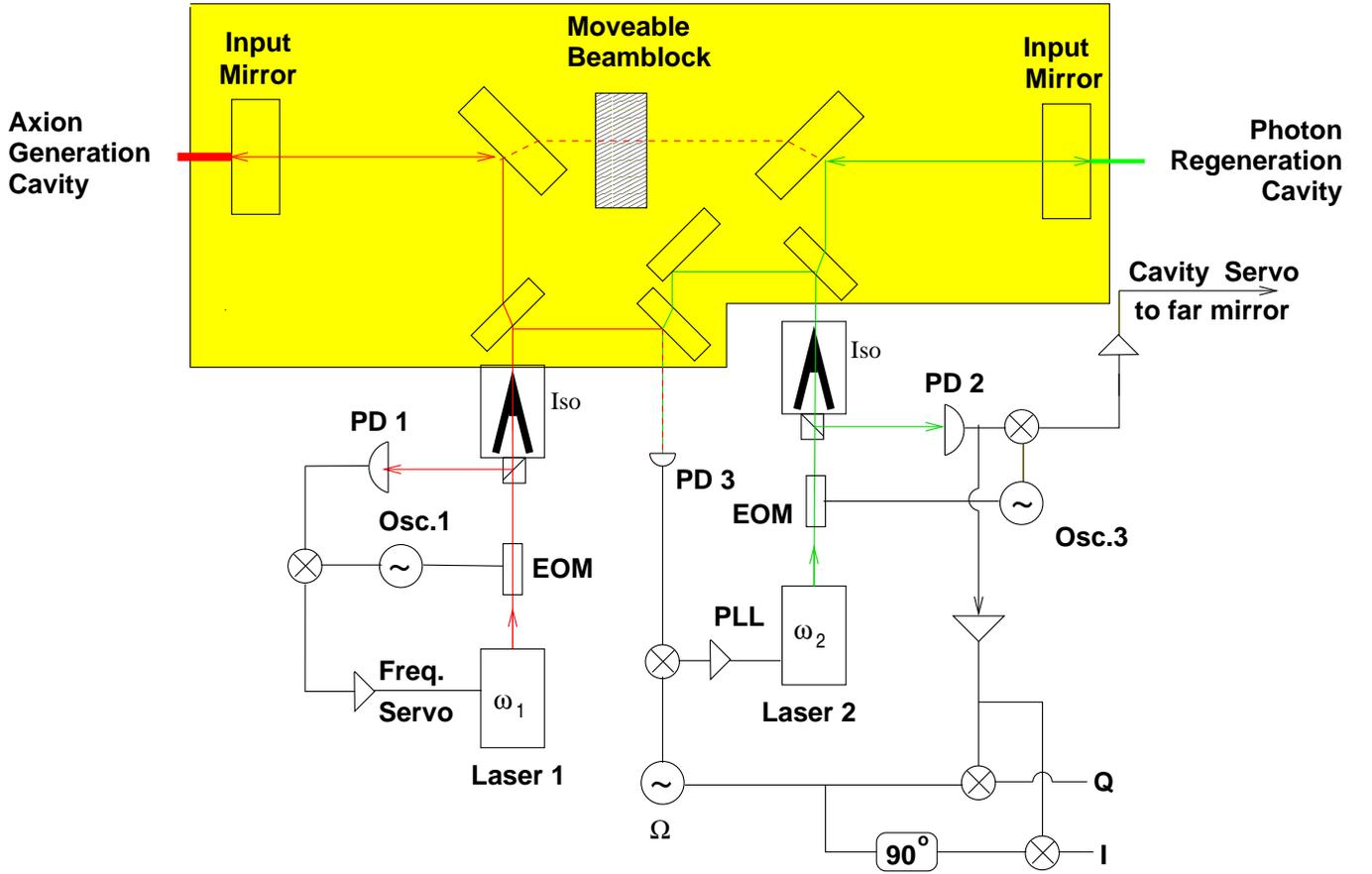} 
\caption{\label{fig:Figure-4.-Schematic}Schematic of the Pound-Drever-Hall
stabilization and locking technique of the two cavities, and the heterodyne
demodulation scheme for detection of the signal. EOM: Electro-optical Modulator PLL: Phase Lock Loop,
Iso: Optical Isolator, PD: Photo detector. The red dashed line indicates
the beam line of the leakage field from the Axion generation cavity.
After removing the beam block, this beam can be used to align both
cavities.}
\end{figure}

Intrinsic to resonantly enhanced photon regeneration is the requirement
that the axion generation and photon regeneration cavities are both on resonance
with the axion generating laser field $E_{in}$. The proposed experimental
setup to achieve these conditions is shown in Figure \ref{fig:Figure-4.-Schematic}.
The laser frequency $\omega_{1}$ of Laser 1 which resonates inside
the axion generation cavity will be stabilized to the eigenfrequency
of the cavity by means of a modulation/demodulation technique commonly
known as the Pound-Drever-Hall technique \cite{PDH}. In this technique,
the phase of the laser field is modulated with an electro-optic modulator
(EOM) before it enters the cavity. This modulation creates a pair
of sidebands offset from the carrier by $\pm$ the modulation frequency. These
sidebands are not resonant inside the cavity and are directly reflected
at the front mirror of the optical cavity while the carrier enters
the cavity. The cavity internal carrier field then transmits back
out through the front mirror, where it is superimposed on the directly
reflected carrier field. If the carrier is on resonance in the optical
cavity, its round-trip phase shift is a multiple of 2$\pi$ and the
leakage field is $\pi$ out of phase with the directly reflected component.
Consequently, the superposition between carrier and sidebands maintains
its phase modulation character, \textit{i.e.}, the generated photocurrent
does not show any modulation at the phase modulation frequency. If
the carrier is slightly off resonance in the optical cavity, its round trip
phase shift is not exactly a multiple of 2$\pi$, and the superposition
between carrier and sidebands can no longer be described as a pure
phase modulation but includes now also some amplitude modulation.
The amplitude modulation is directly proportional to the round trip
phase shift (modulo 2$\pi$) or the frequency offset between the carrier
field and the resonance frequency of the cavity. Moreover, it changes
sign as the cavity passes through resonance. This signal is detected,
amplified, filtered and fed back to the laser's frequency actuators.

Axions regenerated as photons in the photon regeneration cavity will have the same
frequency as the laser photons. Thus, resonant photon regeneration requires
that an eigenfrequency of this cavity must be well within the
linewidth of the resonance of the axion generation cavity. This resonance condition
may be achieved and maintained by means of Laser 2 of frequency
$\omega_{2}$ that is offset from the frequency of Laser 1. 
Light from Laser 2 will first be mixed with
some light picked off from Laser 1; the beat signal will
be demodulated with a tunable RF frequency. The demodulated signal
will then be amplified and filtered before it is fed back to the frequency
actuators of the Laser 2 to maintain a heterodyne phase lock
between the two lasers. The difference frequency is then identical
to the RF frequency $\Omega=\omega_{1}-\omega_{2}$, which will be
set to equal a multiple of the free spectral range of the photon regeneration cavity.
The field from Laser 2 is then injected into the photon regeneration
cavity and one of the eigenfrequencies of this cavity is then
locked to the frequency of Laser 2 using again the Pound-Drever-Hall
scheme described above, the sole difference being that here the feedback
controls the length of the second cavity (through piezo transducers acting on one of the mirrors),
rather than controlling the laser frequency as is done in the axion generation
cavity.

\subsection*{Heterodyne detection of the signal}

The axion field will coherently generate a signal field $E_{S}$ at
laser frequeny $\omega_{1}$ with a phase which depends on the geometric
distance between the two cavities. After leaving the photon regeneration
cavity, this field has the form (see equation \ref{eq:E_S out}):
\begin{equation}
E_{S}=E_{SO}e^{i\omega_{1}t}e^{i\phi}\qquad\phi=k_{a}d\label{eq:E_shete}
\end{equation}
 This field, mode matched to the photon regeneration cavity, will propagate
towards the photodetector which is also used to generate the Pound-Drever-Hall
signal for the cavity stabilization. In addition, the
field from Laser 2 is used as the local oscillator for the signal field. The photodiode
signal is the beat between the two fields, given by: 
\begin{eqnarray}
S&=&\left|E_{SO}e^{i(\omega_{1}t+\phi)}+E_{LO}e^{i\omega_{2}t}\right|^{2}\nonumber \\
&=& E_{LO}^{2}+2E_{LO}E_{SO}\cos\left(\Omega t+\phi\right) \label{eq:S_hete}
\end{eqnarray}
where we have assumed that $E_{SO}\ll E_{LO}$. The limiting noise
source in the detection process will be shot noise. Therefore, it
is convenient to express the signal in terms of the number of photons
in both fields: 
\begin{equation}
S=N_{LO}+S_{I}\cos\Omega t-S_{Q}\sin\Omega t\label{eq:S_N_photon}
\end{equation}
where:
\begin{equation}
S_{I}=2\sqrt{N_{LO}N_{S}}\cos\phi\qquad S_{Q}=2\sqrt{N_{LO}N_{S}}\sin\phi\label{eq:S_IS_Q_N_P}
\end{equation}
are the two quadrature components of the signal.

The shot noise or variance in each quadrature can be calculated from
the number of photons detected: 
\begin{equation}
\sigma_{I}=\sqrt{2\bar{N}}=\sqrt{2N_{LO}}=\sigma_{Q}.\label{eq:SN_ISN_Q}
\end{equation}
The factor $\sqrt{2}$ is a fundamental consequence of the detection
process. The local oscillator beats with the vacuum fluctuations at
frequency $\omega_{0}+\Omega$ and at $\omega_{0}-\Omega$ to generate
a beat signal at $\Omega$. These two contributions are statistically
independent and add quadratically. The signal to shot noise ratio
is then: 
\begin{equation}
\frac{S_{I}}{\sigma_{I}}=\sqrt{2N_{S}}\cos\phi;\qquad\frac{S_{Q}}{\sigma_{Q}}=\sqrt{2N_{S}}\sin\phi.\label{eq:S_I/SN_I}
\end{equation}
If we could adjust the phase to guarantee that $\phi\ll1$ all the
time, our signal to noise in the in-phase quadrature would be $\sqrt{2N_{S}}$.
This phase includes for example the distance traversed by
the axion field, which is not commensurate with the other optical
paths. Therefore, it is currently not clear how to adjust this phase. Instead,
it is likely that our starting phase during
the detection process is arbitrary and that we have to combine both
quadratures to measure the amplitude in the signal: 
\begin{equation}
S_{\Sigma}=\sqrt{S_{I}^{2}+S_{Q}^{2}}=2\sqrt{N_{LO}N_{S}}.\label{eq:S_AMP}
\end{equation}
Similarly, the shot noise in both quadratures adds quadratically:
\begin{equation}
\sigma_{\Sigma}=\sqrt{\sigma_{I}^{2}+\sigma_{Q}^{2}}=2\sqrt{N_{LO}}\label{eq:SN_AMP}
\end{equation}
and the signal to shot noise is: 
\begin{equation}
\frac{S_{\Sigma}}{\sigma_{\Sigma}}=\sqrt{N_{S}},\label{eq:S_SN_AMP}
\end{equation}
where $N_{S}$ is the number of regenerated photons in the signal
field. As expected, to obtain a signal to noise ratio of one requires
one detected photon.

Using equation \ref{prob}, assuming $qL/2\ll1$, and impedance matched
cavities ($T_{1}\approx V$), the number of regenerated photons can
be approximated as follows: 
\begin{eqnarray}
N_{S} &=&\eta^{2}\frac{{\cal F}_{\gamma}}{\pi}\frac{{\cal F}_{a}}{\pi}\frac{1}{16}(gB_{0}L)^{4}N_{in} \nonumber \\
&=&\eta^{2}\frac{{\cal F}_{\gamma}}{\pi}\frac{{\cal F}_{a}}{\pi}\frac{1}{16}(gB_{0}L)^{4}\frac{P_{in}}{\hbar\omega_{0}}\tau \label{eq:N_Slast}
\end{eqnarray}
 where $\tau$ is the measurement time and $P_{in}$ is the power
of the first laser coupled into the axion generation cavity. During
this measurement time the standard deviation in the number of detected
photons increases with $\sqrt{\tau}$. Consequently, the signal to
shot noise increases as $\sqrt{\tau}$.

\subsection{\label{noise}Noise sources and Efficiencies}

So far, this analysis assumes that the phase $\phi$ is constant.
Changes in the detection phase can be understood as phase modulation of
the signal: 
\begin{eqnarray}
\phi(t)&=&<\phi(t)>_{T}+\delta\phi(t)\label{eq:phi_fft} \\
&=&\phi_{0}+\frac{1}{\sqrt{2\pi}}\int_{-\infty}^{\infty}\tilde{\phi}(f)e^{i2\pi ft}df \nonumber \\
e^{i(\omega_{0}t+\phi)}&\approx& e^{i{(\omega}_{0}t+\phi_{0})}\left[1+\frac{i}{\sqrt{2\pi}}\int_{\-\infty}^{\infty}\tilde{\phi}(f)e^{i2\pi ft}df\right] \nonumber 
\end{eqnarray}
 where $\tilde{\phi}(f)$ is the linear spectral density of the phase
fluctuations. This modulation shifts power into other frequency components
and reduce the main signal. Consequently, the amplitude of the phase
modulation has to be sufficiently small. This condition is roughly
equivalent to requesting that the rms-phase fluctuations stay below
about $1\,\mbox{rad}$: 
\begin{equation}
\delta\phi_{rms}=\sqrt{\int_{1/T}^{\infty}\tilde{\phi}^{2}(f)df}<1\,\mbox{rad}
\label{eq:delta_phi_rms}
\end{equation}

Note that this phase is the differential phase between the two laser
fields, which is controlled by a heterodyne phase-locked loop. The reference
signal for this phase-locked loop is also used to demodulate the signal
itself. Sources for potential phase changes are changes in the distance
between the two cavities, differential changes in the distance between
$BS_{1}$ and $BS_{2}$ and between $BS_{1}$ and the axion generation
cavity, and changes in the phase of the reference signal. A satisfactory
solution could be to mount all optical components on a breadboard made from a material with a very low coefficient of thermal expansion such as a Zerodur or ULE and temperature stabilize the entire breadboard to
keep the geometric and optical distances stable. Large variations
in the phase of the reference signal are not expected, as this is
a radio frequency phase and typical cable lengths are shorter than
the wavelength of the RF signal. Length changes of the photon regeneration
cavity will in first order change the phase of both fields by the
same amount. Note that the axion generation cavity is the reference
length of the entire experiment and any changes in it will be tracked by
the laser fields.

Not every photon which reaches the active area of the photodetector
will generate an electron. Some photons will be reflected; others
will excite electron-hole pairs which recombine in the active area.
These effects reduce the efficiency of the photodetector. Typical
quantum efficiencies of commercial photodetectors are on the order
of 80\% but several InGaAs detectors have reached quantum efficiencies
of 95\% \cite{photodet}.

One technical noise source that needs to be considered is the dark
current of the photodetector. This is usually measured in terms of
noise equivalent power (NEP) per root Hz (or equivalently the noise
at the measurement frequency integrated over a measurement time of one second). The beat frequency
is equal to the free spectral range of the cavity which is approximately
3.3 MHz for a 45 m long cavity. Typical photodetectors with sufficient
bandwidth have a noise equivalent power of less than 1 pW/$\sqrt{\mbox{Hz}}$
in the frequency range of interest, and saturate at an input power
of a few mW. Assuming a power of the local laser of $P_{LO}$ =1 mW,
the shot noise would be: 
\begin{equation}
\sigma=15\frac{\mbox{pW}}{\sqrt{\mbox{Hz}}}
\label{eq:sigma_SN}
\end{equation}
 well above the noise of the photodetector.

\section{\label{numeval}Numerical evaluation and discussion}

\begin{figure}
\includegraphics[width=\columnwidth]{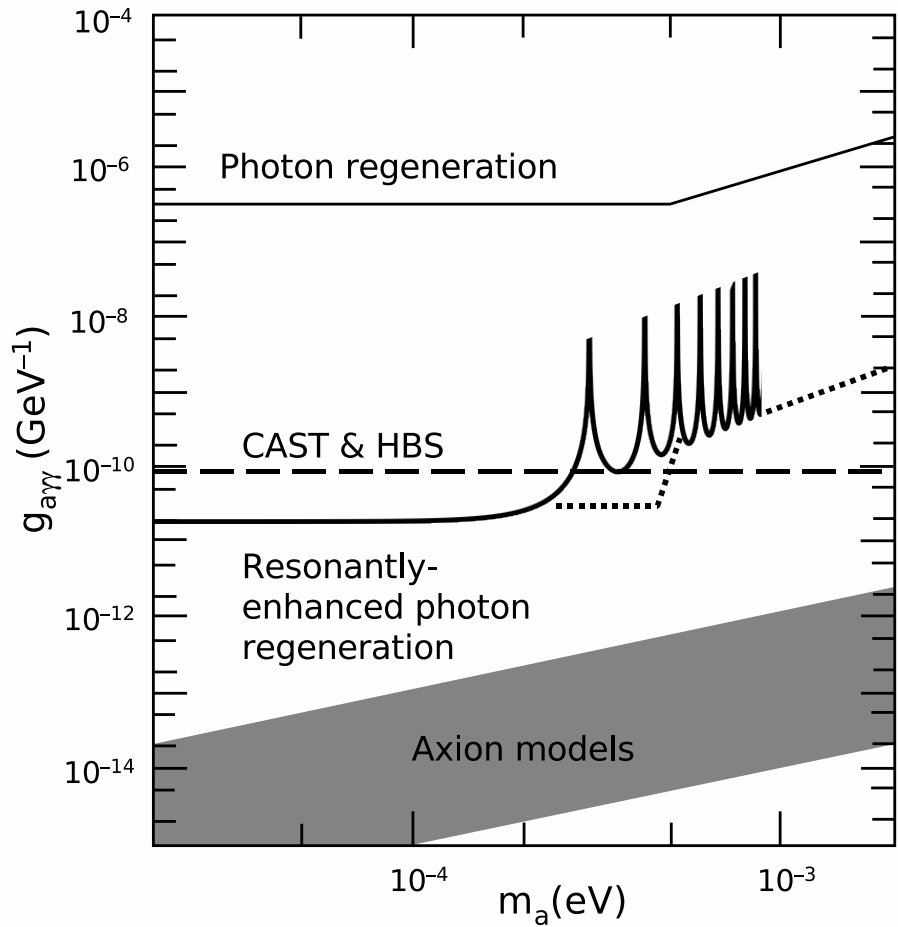} 
\caption{\label{fig:REPR}Current exclusion plot of mass and photon coupling $(m_{a},g_{a\gamma\gamma})$
for the axion, and the 95\% CL exclusion limit for the resonantly
enhanced photon regeneration experiment calculated for the TeV 6+6
configuration and with a cavity finesse of $3.1~\times~10^{5}$. The
solid curve represents the exclusion region with all magnets of the
same polarity; the dotted lines indicate the additional reach in mass
by running in all {}``wiggler'' configurations, as described in
the text. The existing exclusion limits indicated on the plot include the best direct solar axion search
(CAST collaboration) \cite{cast}, the Horizontal Branch Star limit
\cite{RaffeltBook}, and previous laser experiments.\cite{Ruos92,Came93,Robi07,Chou08}}

\end{figure}

For the example of the TeV 6+6 setup described above, an input power
of $10\,\mbox{W}$, a cavity finesse of ${\cal F}\sim\pi\times10^{5}$
($T=10\,\mbox{ppm}=V)$ for both cavities, and a $SNR=1$ we find 
\begin{eqnarray}
g_{a\gamma\gamma}^{min}&=&\frac{2.0\times10^{-11}}{\mbox{GeV}}\left[\frac{0.95}{\eta}\right]\left[\frac{180\,\mbox{Tm}}{BL}\right]\left[\frac{10\,\mbox{ppm}}{T}\right]^{1/2}\nonumber \\
&\times& \left[\frac{10\,\mbox{W}}{P_{in}}\right]^{1/4}\left[\frac{10\,\mbox{days}}{\tau}\right]^{1/4}
\label{eq:g_min_1}
\end{eqnarray}
after 10 days of measurement time. This translates into a 95\% exclusion
limit ($3\sigma$) for axions or generalized pseudoscalars with $g_{a\gamma\gamma}^{min}<2.0\times10^{-11}\mbox{GeV}^{-1}$after
90 days cumulative running, well into territory unexplored by stellar
evolution bounds or direct solar searches. Note that the exclusion sensitivity follows the inverse of $\mbox{sinc}(qL/2)$, for the case of the TeV 6+6 configuration, the first null sensitivity occuring at $2.8\times10^{-4}\mbox{eV}$. The momentum mismatch between a massless photon and a massive axion defines the oscillation length of the process to be $L_{osc}=2\pi/q$.

As pointed out in Ref. \cite{vanB87} however, there is a practical
strategy to extend the mass range upwards if the total magnetic length
$L$ is comprised of a string of $N$ individual identical dipoles
of length $l$. In this case, one may configure the magnet string
as a {}``wiggler'' to cover higher regions of mass, up to values
corresponding to the oscillation length determined by a single dipole,
i.e. $q\sim l^{-1}$. The dotted boundaries in figure 4 depicts how the ``wiggler'' configurations can extend the
mass reach of the exclusion regions; e.g. in the case
of the TeV 6+6 setup, additional running in the combinations of magnet
configurations $\uparrow\uparrow\uparrow\uparrow\uparrow\uparrow$,
$\uparrow\uparrow\uparrow\downarrow\downarrow\downarrow,$, $\uparrow\uparrow\downarrow\downarrow\uparrow\uparrow$
and $\uparrow\downarrow\uparrow\downarrow\uparrow\downarrow$ extend
the mass reach by a factor $\sqrt{6}$ up to $\sim6\times10^{-4}$
eV. 

Clearly the most efficient way to extend the discovery (or exclusion)
potential of the experiment to lower values of $g_{a\gamma\gamma}$ is to make a longer
and/or stronger magnetic string; $g_{a\gamma\gamma}\propto(BL)^{-1}$. While resonant
photon regeneration marks a significant improvement over the simple
experiment, in fact the sensitivity in $g_{a\gamma\gamma}$ still only gains (or loses)
as ${\cal F}^{1/2}$. Thus in our example, the experiment would still
reach a limit of $g_{a\gamma\gamma}=6.2\times10^{-11}$ GeV$^{-1}$ even if the Fabry-Perot
only achieved a finesse of ${\cal F}\simeq30,000$. Assuming $10\,\mbox{ppm}$
intra cavity losses without diffraction, the optimum magnet length
is reached when the aperture causes additional $5\,\mbox{ppm}$ losses.
For a clear aperture of $39\,\mbox{mm}$ diameter, the optimum length
would then be $L_{opt}\approx90\,\mbox{m}$ and, assuming impedance
matched cavities, an intra cavity power of $1\,\mbox{MW}$, and 5T-Tevatron
magnets, $g_{a\gamma\gamma}^{min}\approx8.7\times10^{-12}\mbox{GeV}^{-1}$. This limit
can further be improved using straight Tevatron magnets with a clear
aperture of $48\,\mbox{mm}$ and an optimum length of $\approx140\,\mbox{m}$
to $g_{a\gamma\gamma}^{min}\approx5.8\times10^{-12}\mbox{GeV}^{-1}$.

Thus without future dramatic improvements in superconducting magnet
technology, the reach in sensitivity for resonantly-enhanced photon
regeneration will likely fall short of the axion model band in parameter
space by roughly two to three orders of magnitude at $m_{a}\sim10^{-4}$
eV, and three to four orders of magnitude at $m_{a}\sim10^{-5}$ eV.
Presumably the microwave cavity experiments\cite{Siki83,ARNPS} will
be able to cover the 10 $\mu$eV range with adequate sensitivity
to find or exclude the axion, assuming that axions constitute the
dominant component of the Milky Way halo dark matter. Of course, we
have no \textit{a priori} knowledge of the axion model found in nature
or even if it is to be found at all, and thus one should be open to
surprise once any new region of parameter space becomes accessible.

\begin{acknowledgments}

We thank Muzammil Arain, Tom Carruthers, Aaron Chou, Kem Cook, Bruce
Macintosh, Frank Nezrick, Georg Raffelt, Dan Stancil,  Ray Weiss, and Stan Whitcomb   for useful conversations
during the development of this concept. This work was supported in
part under the auspices of the U.S. Department of Energy under contracts
DE-FG02-97ER41029, and DE-AC52-07NA27344. P.S. gratefully acknowledges
the hospitality of the Aspen Center of Physics while working on this
project.
\end{acknowledgments}



\begin{thebibliography}{10}

\bibitem {Brad03} R. Bradley, John Clarke, Darin Kinion, Leslie J. Rosenberg, Karl van Bibber, Seishi Matsuki, Michael M\"uck, Rev. Mod. Phys. {\bf 75}, 777 (2003). \par

\bibitem{Svrc06} P. Svr\v{c}ek and E. Witten, J. High Energy
Phys. {\bf 0606}, 051 (2006).\par

\bibitem{Siki83} P. Sikivie, Phys. Rev. Lett. {\bf 51} 1415 (1983)

\bibitem{vanB87} K. van Bibber, N.R. Dagdeviren, S.E. Koonin,
A.K. Kerman and H.N. Nelson, Phys. Rev. Lett. {\bf 59}, 759 (1987); for
the case of exactly massless bosons, see A.A. Ansel'm, Yad.\ Fiz.
{\bf 42}, 1480 (1985). \par

\bibitem{Hoogeveen91} Production and detection of light bosons
using optical resonators. F. Hoogeveen, T. Ziegenhagen (Hannover U.).
DESY-90-165, ITP-UH-5-1990, Nov 1990. 28pp. Published in Nucl. Phys.
B {\bf 358}, 3 (1991).\par

\bibitem{Siki07} P. Sikivie, D.B. Tanner, and Karl van Bibber
Phys. Rev. Lett. 98, 172002 (2007). \par

\bibitem{LIGO-NIM} B. Abbott et al.(LIGO
Scientific Collaboration, Nucl. Instrum. Methods A {\bf 517} 154--179
(2004). \par

\bibitem{PQ} R. D. Peccei and H. Quinn, Phys. Rev. Lett.
{\bf 38}, 1440 (1977) and Phys. Rev. D {\bf 16}, 1791 (1977). \par

\bibitem{WandW} S. Weinberg, Phys. Rev. Lett. {\bf 40}, 223 (1978);
F. Wilczek, Phys. Rev. Lett. {\bf 40}, 279 (1978). \par

\bibitem{DFSZ} M. Dine, W. Fischler and M. Srednicki, Phys.
Lett. {\bf 104B}, 199 (1981); A. P. Zhitnitskii, Sov. J. Nucl.
Phys. {\bf 31}, 260 (1980). \par

\bibitem{KSVZ} J. Kim, Phys. Rev. Lett. {\bf 43}, 103
(1979); M. A. Shifman, A. I. Vainshtein and V. I. Zakharov,
Nucl. Phys. B166, 493 (1980). \par

\bibitem{Vac.} L. Abbott and P. Sikivie, Phys. Lett. {\bf 120B},
133 (1983); J. Preskill, M. Wise and F. Wilczek, Phys. Lett. {\bf 120B},
127 (1983); M. Dine and W. Fischler, Phys. Lett. {\bf 120B}, 137 (1983). \par

\bibitem{SN1987a} M. S. Turner, Phys. Rep. {\bf 197}, 67 (1990);
G. G. Raffelt, Phys. Rep. {\bf 198}, 1 (1990). \par

\bibitem{RaffeltBook} G.G. Raffelt, Stars as Laboratories
for Fundamental Physics (University of Chicago Press, Chicago, 1996). \par

\bibitem{ARNPS} S. Asztalos, L. Rosenberg, van K. Bibber,
P. Sikivie, Konstantin Zioutas, Ann. Rev. of Nuc. Part. Sci. {\bf 56} 293-326 (2006). \par

\bibitem{cast} K. Zioutas et al. (CAST collaboration), Phys. Rev. Lett. {\bf 94}, 121301 (2005); E. Arik et al. (CAST collaboration) J. Cosmo. Astropart. Phys. 
(2009). \par

\bibitem{Raff88} G. Raffelt and L. Stodolsky, Phys. Rev.
D {\bf 37}, 1237 (1988). \par

\bibitem{Ruos92}  G. Ruoso, R. Cameron, G. Cantatore, A. C. Melissinos, Y. Semertzidis, H. J. Halama, D. M. Lazarus, A. G. Prodell, F. Nezrick, C. Rizzo and E. Zavattini, Z. Phys. C {\bf 56}, 505 (1992). \par

\bibitem{Came93} R. Cameron, G. Cantatore, A. C. Melissinos, G. Ruoso, and Y. Semertzidis, H. J. Halama, D. M. Lazarus, and A. G. Prodell, F. Nezrick, 
C. Rizzo and E. Zavattini, Phys. Rev. D {\bf 47}, 3707 (1993). \par

\bibitem{Robi07} C. Robilliard, R. Battesti, M. Fouché, J. Mauchain, A.-M. Sautivet, F. Amiranoff, and C. Rizzo, Phys. Rev. Lett. {\bf 99} 190403(4) (2007). \par

\bibitem{Chou08} A. S. Chou, W. Wester, A. Baumbaugh, H. R. Gustafson, Y. Irizarry-Valle, P. O. Mazur, J. H. Steffen, R. Tomlin, X. Yang, and J. Yoo, Phys.
Rev. Lett. {\bf 100} 080402(4) (2008). \par

\bibitem{LIPPS} A. Afanasev, O. K. Baker, K. B. Beard, G. Biallas, J. Boyce, M. Minarni, R. Ramdon, M. Shinn, and P. Slocum, Phys. Rev. Lett. {\bf 101} 120401 (2008) \par

\bibitem{OSQAR} Pierre Pugnat, Lionel Duvillaret, Remy Jost, Guy Vitrant, Daniele Romanini, Andrzej Siemko, Rafik Ballou, Bernard Barbara, Michael Finger, Miroslav Finger, Jan Ho{\v s}ek, Miroslav Kr{\'a}l, Krzysztof A. Meissner, Miroslav {\v S}ulc, and Josef Zicha, Phys. Rev. D {\bf 78}, 092003 (2008)\par

\bibitem{Ringwald:2006rf} Several other photon regeneration
experiments either in preparation or proposed are found in A.~Ringwald,
arXiv:hep-ph/0612127. \par

\bibitem{HelloVinet} J.-Y. Vinet, P. Hello, C.N. Man, A. Brillet, J. Phys. (Paris) I {\bf 2} 1287-1303 (1992), P. Hello, J.-Y. Vinet, J. Phys. France {\bf 51} 1267 (1990),  P. Hello, J.-Y. Vinet, J. Phys. France {\bf 51} 2243 (1990)  \par

\bibitem{Lawrence} Ryan Lawrence, {\it Active Wavefront Correction in Laser Interferometric Gravitational Wave Detectors}, Ph.D. thesis, MIT (2003).


\bibitem{Siegman} A.E. Siegman, {\it Lasers}, University Science Books, Sausalito, CA (1984).\par

\bibitem{PDH} R.W.P. Drever, J.L. Hall, F.V. Kowalski, J.
Hough, G.M. Ford, A.J. Munley, and H. Ward, Appl. Phys. B {\bf 31}, 97 (1983);
Eric D. Black, Am. J. Phys. {\bf 69}, 79 (2001). \par

\bibitem{photodet} ETX-500 data sheet at http://www.jdsu.com/product-literature/etx500\_ds\_cc\_ae.pdf. \par


\textbf{
}

\end{thebibliography}
\end{document}